\documentclass{aa}

\usepackage{graphicx}
\usepackage[varg]{txfonts}
\usepackage{amsmath}	
\usepackage{bigstrut,multirow,rotating}
\usepackage{float}

\usepackage[colorlinks=true,linkcolor=blue, citecolor=blue, filecolor=blue, urlcolor=blue]{hyperref}

\begin{document}
\title{Forecast of strongly lensed supernovae rates in the China Space Station Telescope surveys}

\author{Jiang Dong\inst{1,2} \and 
        Yiping Shu\inst{1}\and 
        Guoliang Li\inst{1} \and 
        Xinzhong Er\inst{3} \and 
        Bin Hu\inst{4,5} \and 
        Youhua Xu\inst{6}}

\institute{Purple Mountain Observatory,Chinese Academy of Sciences, Nanjing 210023, People’s Republic of China\\
           \email{yiping.shu@pmo.ac.cn}
           \and
           School of Astronomy and Space Science, University of Science and Technology of China, Hefei, Anhui, 230026, People’s Republic of China
           \and
           South-Western Institute for Astronomy Research, Yunnan University, Kunming, Yunnan 650504,People's Republic of China
           \and
           Institute for Frontier in Astronomy and Astrophysics, Beijing Normal University, Beijing 102206, People's Republic of China
           \and
           Department of Astronomy, Beijing Normal University, Beijing 100875, People's Republic of China
           \and
           National Astronomical Observatories, Beijing, 100101, People's Republic of China}
   
\date{Received ...; accepted ...}

\abstract{
Strong gravitationally lensed supernovae (SNe) are a powerful probe for cosmology and stellar physics. The relative time delays between lensed SN images provide an independent way of measuring a fundamental cosmological parameter --- the Hubble constant ---, the value of which is currently under debate. The time delays also serve as a ``time machine'', offering a unique opportunity to capture the extremely early phase of the SN explosion, which can be used to constrain the SN progenitor and explosion mechanism. Although there are only a handful of strongly lensed SN discoveries so far, which greatly hinders scientific applications, the sample size is expected to grow substantially with next-generation surveys. In this work, we investigate the capability of detecting strongly lensed SNe with the China Space Station Telescope (CSST), a two-meter space telescope to be launched around 2026. Through Monte Carlo simulations, we predict that CSST can detect 1008.53 and 51.78 strongly lensed SNe from its Wide Field Survey (WFS, covering 17,500 deg$^2$) and Deep Field Survey (DFS, covering 400 deg$^2$) over the course of ten years. In both surveys, about 35\% of the events involve Type Ia SNe as the background sources. Our results suggest that the WFS and DFS of CSST, although not designed or optimized for discovering transients, can still make a great contribution to the strongly lensed SNe studies. 
}


\titlerunning{Strongly lensed SNe forecast for the CSST}

\maketitle
%
%________________________________________________________________

\section{Introduction}

In a strong gravitationally lensed supernova (SN) system, light traveling along multiple paths is deflected by the intervening lensing object and forms multiple images in the plane of the observer \citep[e.g.,][]{Einstein1936, Zwicky1937}. Because of differences in the geometric length and experienced gravitational potential, the arrival time of the multiple images can exhibit relative delays. Strong-lensing time delays can be used to constrain the current expansion rate of the Universe --- the Hubble constant $H_0$ \citep[e.g.,][]{Refsdal1964} ---, the determination of which is currently a burning issue. There is a more than 5$\sigma$ discrepancy on the $H_0$ measurements from two well-established cosmological probes: the local distance ladder \citep[e.g.,][]{Riess2022} and the cosmic microwave background \citep[CMB, e.g.,][]{2020A&A...641A...6P}, which is known as the Hubble tension. Measuring $H_0$ with other independent approaches is crucial in order to better understand the causes of this tension. The time-delay cosmography approach has been successfully validated with strongly lensed variables such as quasars and SNe \citep[e.g.,][]{Oguri2019,Liao2022,Birrer2022a,Treu2022,Treu2023,Suyu2024}. In particular, a 2.4\% precision on $H_0$ has been achieved from analyzing six strongly lensed quasars \citep{Wong2020}. Compared to quasars, strongly lensed SNe confer several advantages for carrying out the time-delay cosmography: (1) the light curves of SNe are dominated by a single, strong peak in the rest-frame UV/optical and are much better understood and characterized \citep[e.g.,][]{Doggett1985,Woosley2007,Kasen2009,Sanders2015,Morozova2017}; (2) the timescale of SN light curves is substantially shorter and therefore requires far fewer observing resources; (3) the early-phase color curves of SNe are found to be almost immune to microlensing \citep[e.g.,][]{Goldstein2018,Huber2021}, which would otherwise complicate the time-delay determination; (4) the standard candle nature of certain types of SNe can be used to independently constrain the lensing magnifications, which helps to break the mass--sheet degeneracy in the lensing modeling \citep[e.g.,][]{Falco1985,Oguri2003,Schneider2013,Schneider2014}; (5) lensed images of the SN host galaxy can be better detected and modeled before the SN explodes or after it fades away, which has been shown to yield tighter constraints on $H_0$ \citep[e.g.,][]{Oguri2019,Ding2021}.

Furthermore, the strong lensing time delays can be used to study the early phases of SN explosions. It is well recognized that the early-phase SN properties (typically within 10 rest-frame days) are crucial for determining its progenitor, environment, and explosion characteristics \citep[e.g.,][]{Kasen2010,Bulla2020,Li2023a}. However, as SN explosions are generally unpredictable, such early-phase observations are mostly obtained by chance and therefore remain scarce. When an SN is strongly lensed, we are potentially able to predict, after detecting the leading images, where and when the trailing images will appear \citep[e.g.,][]{Treu2016,Kelly2016}. Follow-up observations can thus be scheduled to reveal the whole SN explosion process. 

Despite the great scientific capabilities mentioned above, strongly lensed SNe are intrinsically very rare.
First, strong lensing requires an exceptionally close alignment between the observer, the lensing galaxy, and the background source, and so the typical occurrence rate of galaxy-scale strong lenses is $\sim 10^{-4}-10^{-3}$ \citep[e.g.,][]{Marshall2009, Collett2015,Oguri2018, Cao2023}. Second, the typical SN event rate is found to be a few per galaxy per century \citep{Diehl2006,Adams2013,Shu2018}. So far, only nine strongly lensed SNe have been reported: ``PS1-10afx'' \citep{Quimby2013a}, ``SN Refsdal'' \citep{Kelly2015}, ``iPTF16geu'' \citep{Goobar2017}, ``SN Requiem'' \citep{Rodney2021}, ``AT 2022riv'' \citep{Kelly2022}, ``C22'' \citep{Chen2022}, ``SN Zwicky'' \citep{Goobar2023}, ``SN H0pe'' \citep{Frye2024}, and ``SN Encore'' \citep{Pierel2024}.  Among these, PS1-10afx, SN Zwicky, and iPTF16geu are galaxy-scale events, and the rest are group- or cluster-scale events. In particular, SN Requiem and SN Encore are from the same host galaxy.

A large number of strongly lensed SNe are expected to be discovered by upcoming wide field imaging surveys, such as the China Space Station Telescope \citep[CSST,][]{Zhan2021}, the \textit{Vera Rubin} Observatory Legacy Survey of Space and Time \citep[\textit{Rubin} LSST,][]{Ivezi2019}, the Roman Space Telescope \citep{Spergel2015}, and the Wide Field Survey Telescope \citep[WFST,][]{2023SCPMA..6609512W}. Dedicated estimations of strongly lensed SNe rates from several surveys have already been reported \citep[e.g.,][]{Oguri2010, Goldstein2019, Pierel2021}. 

In the present work, we aim to quantify the rates and properties of strongly lensed SNe that can be detected by the CSST, a flagship 2m space telescope to be launched around 2026. The paper is organized as follows. In Sect. \ref{sec:s2}, we outline how a mock catalog of strongly lensed SNe is generated using Monte Carlo simulations. In Sect. \ref{sec:s3}, we estimate the number of strongly lensed SNe detectable by the CSST under a nominal survey strategy. Discussions and conclusions are presented in Sect. \ref{sec:s4} and Sect. \ref{sec:s5}. Throughout this paper, we adopt a flat $\Lambda$CDM cosmology with $\Omega_{\Lambda}=0.74$, $\Omega_{m}=0.26,$ and $H_{0}$=72 km s$^{-1}$ Mpc$^{-1}$ \citep{Oguri2010}.

\section{Generating the mock catalog}
\label{sec:s2}

This section describes how we populate the lensing galaxies and background SNe separately and generate a mock catalog of strongly lensed SNe across the full 4$\pi$ sky using Monte Carlo simulations. 

\subsection{Lensing galaxies}
\label{sec:s2.1}

%For the lensing galaxy population, we only consider elliptical galaxies, because they are believed to dominate the total lensing cross-sections \citep[e.g.,][]{Oguri2006,2007MNRAS.379.1195M}. 
We assume that the two-dimensional projected mass distribution of the lensing galaxies follows the singular isothermal ellipsoid (SIE) profile \citep{Kormann1994} with convergence $\kappa$ given by:
 \begin{equation}
	\kappa \left( x,y \right) =\frac{\theta _{\mathrm{Ein}}}{2}
	\frac{\lambda \left( e \right)}{\sqrt{\left( 1-e \right) ^{-1}x^2+\left( 1-e \right) y^2}},
 \end{equation}
 \begin{equation}
	\theta _{\mathrm{Ein}}=4\pi \left( \frac{\sigma}{c} \right) ^2\frac{D_{ls}}{D_s}
	\label{eq:e2},
 \end{equation}
where $\theta_{\mathrm{Ein}}$ is the Einstein radius, $\sigma$ is the line-of-sight velocity dispersion of the lensing galaxy, $e$ is the ellipticity of the projected mass distribution, and $D_{ls}$ and $D_{s}$ are the angular diameter distances from the lensing galaxy and the observer to the background source, respectively. The factor $\lambda (e)$ is the so-called dynamical normalization parameter, which is related to the three-dimensional shape of the lensing galaxy mass distribution \citep[e.g.,][]{Chae2003}. Following \citet{Oguri2010}, we consider an equal number of oblate and prolate galaxies in the lensing galaxy population, and adopt the average value of $\lambda (e)$ for each oblate and prolate case\footnote{The average values are 1.237 and 0.923 for oblate and prolate cases. These are obtained by integrating from $e=0$ to $e=0.9$ for oblate cases (to match the adopted ellipticity distribution) and from $e=0$ to $e=0.7$ for prolate cases (because $\lambda$ diverges beyond $e=0.7$ for prolate cases).}. We further assume that the ellipticity follows a truncated normal distribution with a mean of 0.3 and a dispersion of 0.16, truncated at $e=0$ and $e=0.9$ \citep{Oguri2008}. 

The overall abundance and velocity dispersion distribution of the potential lensing galaxies are given by the velocity dispersion function (VDF) in the form of 
\begin{equation}
\phi \left( \sigma, z_l \right) =\phi _* (z_l) \left( \frac{\sigma}{\sigma_*(z_l)} \right) ^{a}
\exp \left[ -\left( \frac{\sigma}{\sigma_*(z_l)} \right) ^{b} \right] \frac{1}{\sigma},
\end{equation}
where $\phi (\sigma, z_l)$ represents the number of potential lensing galaxies at redshift $z_l$ per unit velocity dispersion per unit comoving volume. 
As suggested by \citet{Geng2021} and \citet{Yue2022}, we adopt the following values and redshift dependence for parameters $\phi_*$, $\sigma_*$, $a$, and $b$: $\phi_*=6.92\times10^{-3}(1+z_l)^{-1.18}~\mathrm{Mpc^{-3}}$, 
$\sigma_*=172.2\times(1+z_l)^{0.18}~\mathrm{km~s^{-1}}$, $a=-0.15$, and $b=2.35$. This redshift-dependence VDF is found to be consistent with the observed VDFs out to $z_l \sim 1.5$ \citep[e.g.,][]{Bezanson2012, Hasan2019,Yue2022}. We note that those observed VDFs are constructed using both quiescent and star-forming galaxies, and so the potential lensing galaxies simulated here also include both types of galaxies. In this work, we only consider lensing galaxies in the redshift range from 0 to 2 and velocity dispersion range from 100 km s$^{-1}$ to 450 km s$^{-1}$. The total number of the potential lensing galaxies under consideration is therefore $\approx 8.6 \times 10^8$ across the full sky. For simplicity, we assume the lensing galaxies are uniformly distributed on the sky. 

To account for additional lensing effects from the lens environment and line-of-sight structures, we added an external shear component in the lens model \citep{Kochanek1991,Keeton1997,Witt1997}. The lensing potential of the external shear component is given by:
\begin{equation}
	V \left( x,y \right) =\frac{\gamma}{2}\left( x^2-y^2 \right) \cos 2\theta _{\gamma}-\gamma x y \sin 2 \theta _{\gamma},
\end{equation}
where $\gamma$ and $\theta_{\gamma}$ are the strength and orientation of the external shear. We assume the strength $\gamma$ follows a log-normal distribution with a mean of -1.301 (i.e., $\langle \log_{\rm 10} \gamma \rangle $ = -1.301) and dispersion of 0.2 dex, which are motivated by N-body simulations and semi-analytic models of galaxy formation \citep{Holder2003}. The orientation $\theta_{\gamma}$ is drawn randomly from 0 to 360 degrees. 

\subsection{Background supernovae}
\label{sec:s2.2}

In this work, we consider seven types of SNe, namely normal Ia, 91bg, 91T, Ib/c, IIP, IIL, and IIn. Normal Ia, 91bg, and 91T are three subtypes of Type Ia SNe (SNe Ia) and the rest are subtypes of core-collapse supernovae (CCSNe). SNe Ia originate from the thermonuclear explosions in a binary system, whereas CCSNe originate from the collapse of massive stars.

The SNe Ia volumetric rate is estimated as the convolution of the cosmic star formation history (SFH) and the delay-time distribution  \citep{Dahlen1999,Oguri2010}:
\begin{equation}
	n_{\mathrm{Ia}}=\eta k_{\mathrm{Ia}}\frac{\int^{t(z_{\rm SN})}_{0}{\psi[z(t-t_d)]f(t_d)}dt_d}{\int_{0}^{t(z=0)}{f(t_d)dt_d}} ~\mathrm{yr^{-1}}~\mathrm{Mpc^{-3}},
\end{equation}
where $\psi (z)$ is the SFH, $f (t_d)$ is the delay-time distribution, $\eta$ represents the proportion of progenitors that successfully explode as SNe Ia, and $k_\mathrm{Ia}$ is the number of SNe Ia progenitors per unit mass formed. We adopt the SFH from \citet{Madau2014}, which is obtained by fitting the UV and IR observational data,
 \begin{equation}
	\psi(z)=0.015\frac{(1+z)^{2.7}}{1+[(1+z)/2.9]^{5.6}}~\mathrm{M_{\odot}}~\mathrm{yr^{-1}}~\mathrm{Mpc^{-3}}.
 \end{equation}
For the delay time distribution $f (t_d)$, we adopt the form from \citet{Maoz2012}:
\begin{equation}
    f(t_d) \propto t_d^{-1.07}.
\end{equation}
Parameter $\eta$ is set to a canonical value of 0.04 \citep[e.g.,][]{Hopkins2006,Maoz2012}. 
The parameter $k_\mathrm{Ia} = 0.021 ~ \mathrm{M_{\odot}^{-1}}$ is computed from
\begin{equation}
    k_{\mathrm{Ia}}=\frac{\int_{M_{1}}^{M_{2}}{\varphi \left( M \right) dM}}{\int_{M_{\min}}^{M_{\max}}{M\varphi \left( M \right) dM}},
    \label{eq:e8}
\end{equation}
where $\varphi (M)$ is the stellar initial mass function (IMF) from \citet{Salpeter55}, $M_{\min}=0.1~\mathrm{M_{\odot}}$ and $M_{\max}=125~\mathrm{M_{\odot}}$ are the cutoff masses of the adopted IMF, and $M_1= 3~\mathrm{M_{\odot}}$ and $M_2= 8~\mathrm{M_{\odot}}$ define the mass range of SNe Ia progenitors.

The CCSNe volumetric rate is directly connected to the SFH as \citep{Dahlen1999,Oguri2010}:
\begin{equation}
	n_{\mathrm{cc}}(z_{\rm SN})=k_{\mathrm{cc}}\psi(z_{\rm SN}),
\end{equation}
where $k_{\mathrm{cc}} = 0.0070 ~ \mathrm{M_{\odot}^{-1}}$ can also be computed from Eq. \eqref{eq:e8}, but the mass range of CCSNe progenitors is from $M_1=8~\mathrm{M_{\odot}}$ to $M_2=50~\mathrm{M_{\odot}}$.

\begin{figure}
    \centering
	\includegraphics[width=0.9\columnwidth]{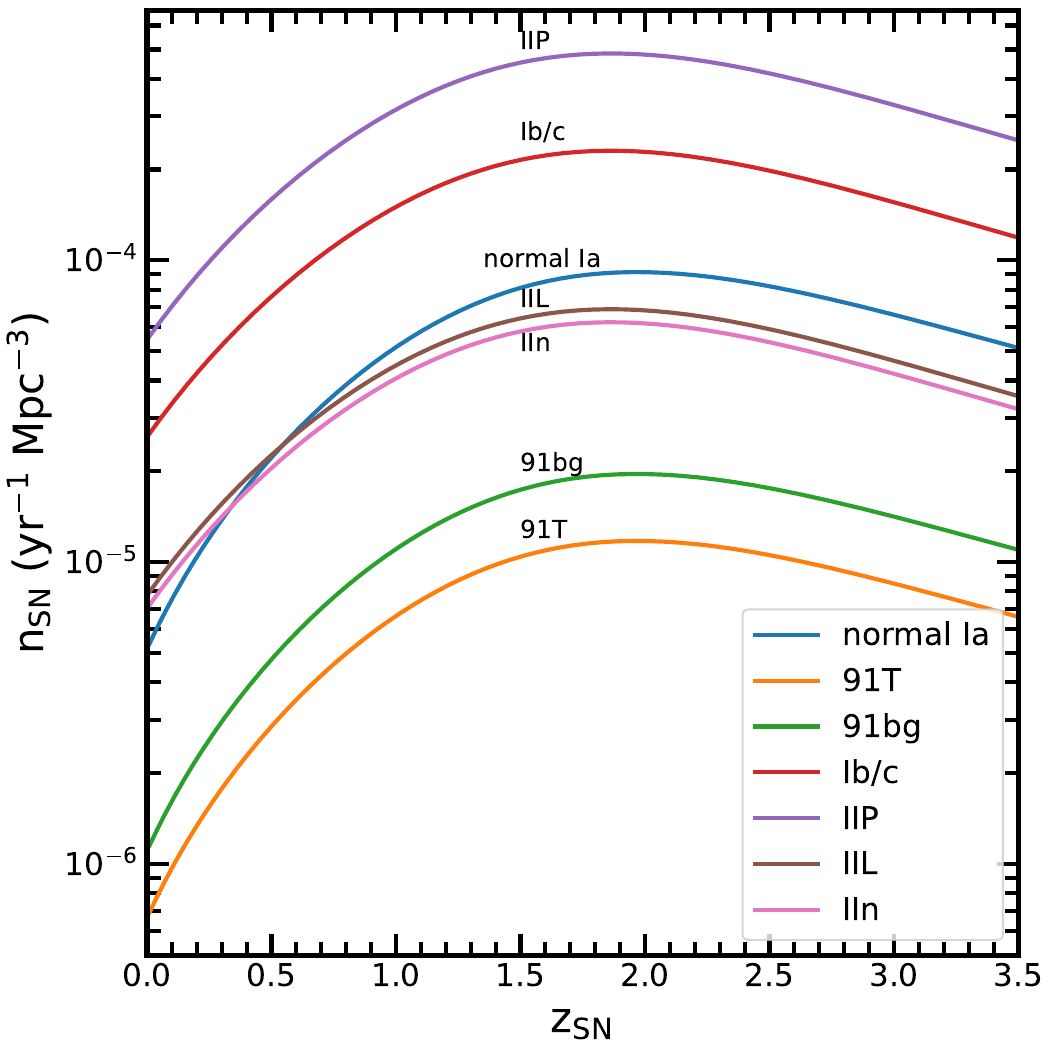}
    \caption{Redshift evolution of the adopted volumetric rates of the seven SN subtypes: normal Ia, 91T, 91bg, Ib/c, IIP, IIL, IIn. }
    \label{fig:sn_rate}
\end{figure}

Following \citet{Li2011}, we assume the relative fractions of the three SNe Ia subtypes and the four CCSNe subtypes in a volume-limited sample are as follows :

\begin{align}
    n_{\mathrm{normal}} &= 0.70~n_{\mathrm{Ia}}, \nonumber \\
    n_{\mathrm{91bg}} &= 0.152~n_{\mathrm{Ia}}, \nonumber \\
    n_{\mathrm{91T}} &= 0.094~n_{\mathrm{Ia}}, \nonumber \\
    n_{\mathrm{Ib/c}} &= 0.246~n_{\mathrm{cc}}, \nonumber \\
    n_{\mathrm{IIP}} &= 0.526~n_{\mathrm{cc}}, \nonumber \\
    n_{\mathrm{IIL}} &= 0.073~n_{\mathrm{cc}}, \nonumber \\
    n_{\mathrm{IIn}} &= 0.064~n_{\mathrm{cc}}. \nonumber
\end{align}
We note that the sum of the relative fractions of either SNe Ia subtypes or CCSNe subtypes is less than unity because some of the SNe subtypes included in \citet{Li2011} are not considered here. The redshift evolution of the event rates of the seven considered SN subtypes is shown in Fig. \ref{fig:sn_rate}.

\begin{table}
	\caption{$R$-band luminosity functions of the seven SN subtypes \citep[from][]{Li2011} and the adopted SN spectral templates (corresponding to source names defined in the \texttt{SNCosmo} package).
     }
	\label{tab:LF}
	\centering
       \scalebox{1}{
	\begin{tabular}{lcccc}
		\hline
		SN Type&$M_{\mathrm{SN}}$&$\sigma_{\mathrm{SN}}$&Spectral Templates&References\\
		\hline
        normal Ia&-18.67&0.51&nugent-sn1a&1\\
		91bg&-17.55&0.53&nugent-sn91bg&1\\
		91T&-19.15&0.52&nugent-sn91t&2\\
		Ib/c&-16.09&1.24&nugent-sn1bc&3\\
		IIP&-15.66&1.23&nugent-sn2p&4\\
		IIL&-17.44&0.64&nugent-sn2l&4\\
		IIn&-16.86&1.61&nugent-sn2n&4\\
		\hline
	\end{tabular}
            }
        \tablebib{
         (1)~\citet{Nugent2002}; (2)~\citet{Stern2004}; (3)~\citet{Levan2005}; (4)~\citet{Gilliland1999}.
        }
\end{table}

We assume that the peak luminosity distributions of SNe follow a normal distribution \citep{Oda2005,Oguri2010} parameterized in the following form:
\begin{equation}
	\phi_{\mathrm{SN}} (M) =\frac{n_{\mathrm{SN}}}{1+z_{\rm SN}}\frac{1}{\sqrt{2\pi}\sigma_{\mathrm{SN}}}\mathrm{exp}[-\frac{(M-M_{\mathrm{SN}})^2}{2\sigma_{\mathrm{SN}}^2}]
\end{equation}
where $\phi_{\mathrm{SN}} (M)$ represents the number of background SNe per unit peak absolute magnitude per unit comoving volume. The $(1+z)^{-1}$ factor is introduced to convert the rest-frame SN rates to the observer-frame rates. 
The mean and dispersion of the $R$-band luminosity function (LF) of different SN subtypes in a volume-limited sample were given in \citet{Li2011} and are summarized in Table \ref{tab:LF}. We note that these LFs have been corrected for the Galactic extinction but not for the host-galaxy extinction. As a result, by adopting these LFs, the host-galaxy extinction has been implicitly included. The specific spectral templates used to generate the light curves for each SN subtype are also given in Table \ref{tab:LF}. For simplicity, we do not take into account the diversities in the spectral time series for individual subtypes. 

\begin{figure*}
    \centering
	\includegraphics[width=0.9\columnwidth]{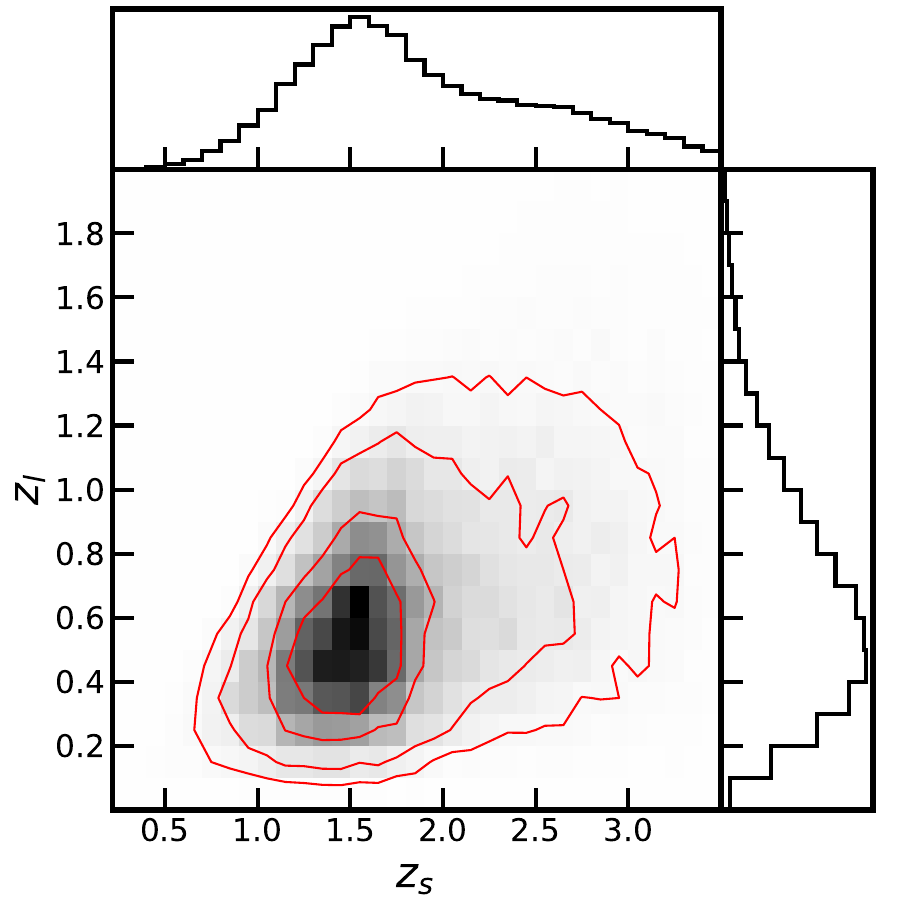}
    \qquad
    \includegraphics[width=0.9\columnwidth]{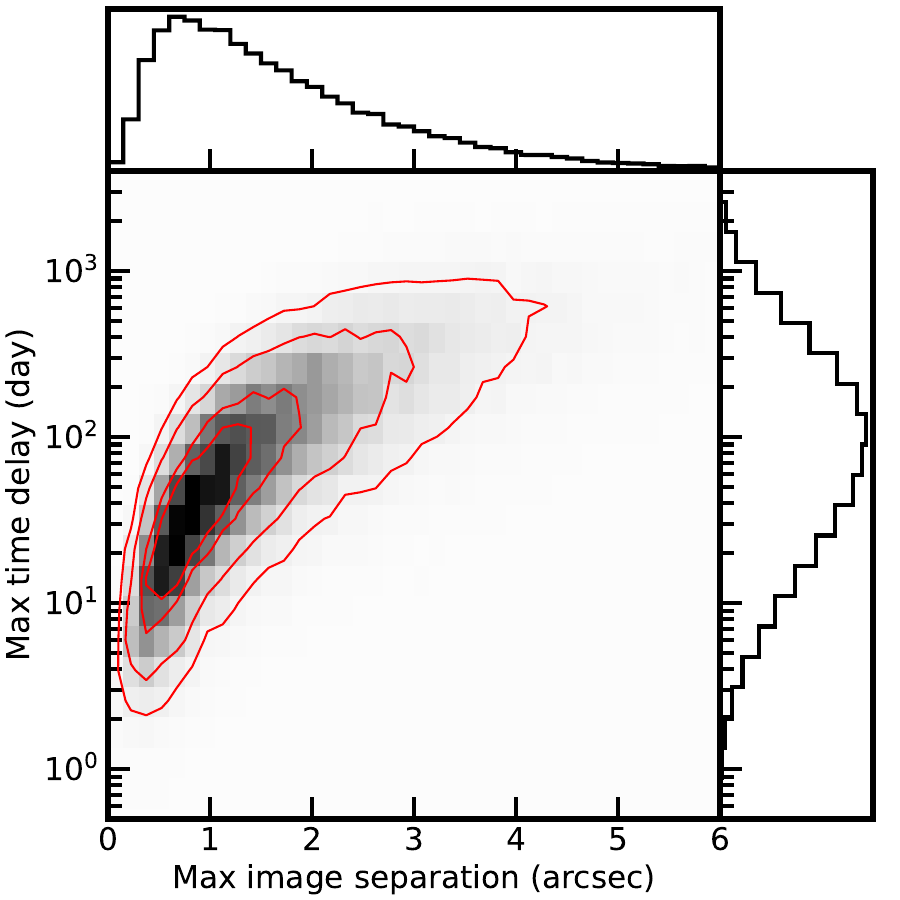}
    \caption{ Selected properties of strongly lensed SNe from our mock catalog. Left: Two-dimensional distributions of the lens and source (i.e., SN) redshifts. Right: Two-dimensional distributions of the maximum image separation and the maximum relative time delay. The red contours enclose 20\%, 35\%, 65\% and 85\% of the full sample. }
    \label{fig:properties of catalog}
\end{figure*}
In this work, we only consider background SNe in the redshift range from 0 to 3.5 and with peak $i$-band apparent magnitude brighter than or equal to 31.5 mag. The total number of SNe (of all subtypes) under consideration is $\approx 1.8 \times 10^8$ per year across the full sky, or $\approx 4348$ per square degree per year. Again, for simplicity, we assume the background SNe distribute uniformly on the sky. 

\begin{figure}
    \centering
	\includegraphics[width=0.9\columnwidth]{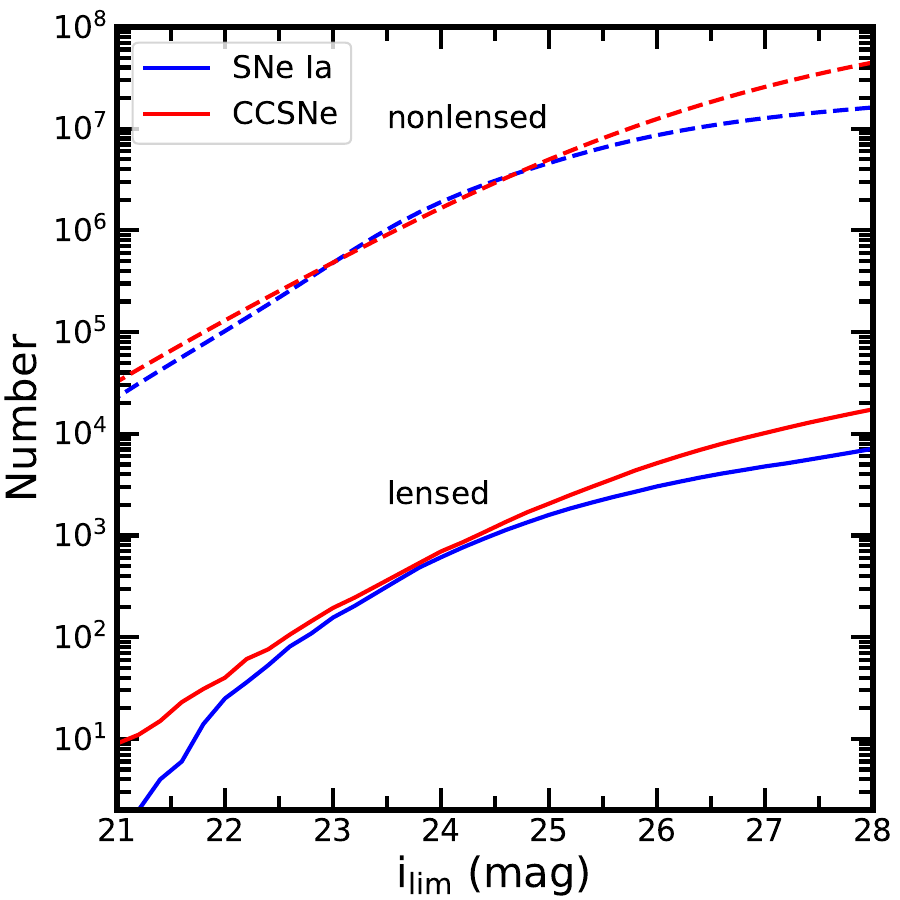}
    \caption{Detectable nonlensed and lensed SNe per year across the full sky as a function of $i$-band limiting magnitude. The dashed line represents the number of nonlensed SNe for SNe Ia and CCSNe, respectively. The solid line represents the number of lensed SNe for the SNe Ia and CCSNe, respectively. }
    \label{fig:non_lensed}
\end{figure}

\subsection{The mock catalog}
\label{sec:s2.3}

After generating a sample of potential lensing galaxies (Sect. \ref{sec:s2.1}) and a sample of background SNe (Sect. \ref{sec:s2.2}), we mainly follow the methodology in \citet{Oguri2010} to generate a mock catalog of strongly lensed SNe using Monte Carlo simulations. Our procedures are summarized as follows. For every lensing galaxy at $z_l$, we randomly select 4348 SNe from the background SN sample, which represent the SNe located within the light cone of one square degree centered on the chosen lensing galaxy. Their projected positions on this one square degree area are randomly assigned. For any SN that is at a redshift higher than $z_l$ and is located closer than $8\theta_{\mathrm{Ein,max}}$\footnote{$\theta_{\mathrm{Ein,max}}$ can be calculated from Eq. \eqref{eq:e2} by setting $D_{ls}/D_{s}$ to one.} from the lensing galaxy in projection, we solve the lens equation using {\tt lfit\_gui} developed in \citet{Shu2016}. If the SN is strongly lensed (i.e., forming multiple images), this system is kept in our mock catalog, and the properties of the SN, its multiple images, and the corresponding lensing galaxy are also saved in the mock catalog. The final mock catalog contains $\approx 6.4 \times 10^4$ systems, which correspond to the expected number of strongly lensed SN events per year across the full sky. Our mock catalog is available as a FITS file at Zenodo \footnote{\url{https://zenodo.org/records/12739548}} and the National Astronomical Data Center \footnote{\url{https://nadc.china-vo.org/res/r101465/}}, and descriptions of all the columns in the FITS file are provided in Table \ref{tab:catalog}.

In our mock catalog, 95.91\% of sources are two-image systems (i.e., doubles), 4.02\% are four-image systems (i.e., quads), and 0.07\% are three-image systems (i.e., naked cusps). Splitting according to SN subtype, the sources in our mock catalog consist of 25.42\% normal Ia, 1.18\% 91bg, 1.35\% 91T, 20.20\% Ib/c, 27.12\% IIP, 6.06\% IIL, and 18.67\% IIn. Interestingly, there are 29 (0.04\%) double-source plane lenses in our mock catalog, which are defined as two different SNe strongly lensed by the same lensing galaxy. As a side note, it was recently discovered that two different SNe (from the same host galaxy) are strongly lensed by the same group of galaxies, namely SN Requiem \citep{Rodney2021} and SN Encore \citep{Pierel2024}. %It has been shown that multiple plane lenses can be used to constrain cosmological parameters \citep[e.g.,][]{Collett2014, Linder2016, Collett2020}. 

Figure \ref{fig:properties of catalog} visualizes the distributions of the mock catalog in terms of lens redshift, source redshift, maximum image separation, and maximum relative time delay. We find that the mean and median lens redshifts are 0.69 and 0.63, and the mean and median source (i.e., SN) redshifts are 1.90 and 1.77. The maximum image separations start from 0\farcs026 and 50\% of the mock strong-lens systems have image separations greater than 1\farcs46. The maximum time delays reach as long as $\sim 1000$ days, and the mean and median are 179 days and 79 days. 

Figure \ref{fig:non_lensed} shows the event rates of detectable nonlensed and lensed SNe as a function of $i$-band limiting magnitude (5$\sigma$, point sources). For this estimation, we define a lensed SN as detectable when the peak $i$-band magnitude of its first arrival image is above the limiting magnitude. It is worth pointing out that the $i$-band limiting magnitude used in \citet{Oguri2010} corresponds to a 10$\sigma$ detection limit for point sources. Our simulations suggest that the galaxy-scale strong-lensing rate for SNe is $\sim10^{-3.5}$, consistent with previous results \citep[e.g.,][]{Oguri2010}. We also find that the lensing rate does not vary significantly with the $i$-band limiting magnitude (within the magnitude range explored).

\begin{table*}
  \centering
  \label{tab:band_parameter}
  \caption{Key specifications of the CSST imaging observations. The numbers in parentheses represent the single-visit limiting magnitudes \citep{Zhan2021}.}
    \scalebox{0.95}{
    \begin{tabular}{lccccccccc}
    \hline
    \hline
    Program & Area (deg$^2$)& Exposure time & \multicolumn{7}{c}{$5\sigma$ limiting magnitude for point sources (AB mag)} \bigstrut\\
    \hline
      &   &   & NUV\tablefootmark{a} & $u$ & $g$ & $r$ & $i$ & $z$ & $y$\tablefootmark{a} \bigstrut\\
\cline{4-10}    WFS & 17,500 & $2 \times 150$ s & 25.4 (24.6) & 25.4 (25.0) & 26.3 (25.9) & 26.0 (25.6) & 25.9 (25.5) & 25.2 (24.8) & 24.4 (23.6) \bigstrut[t]\\
                DFS & 400 & $8 \times 250$ s & 26.7 (25.2) & 26.7 (25.6) & 27.5 (26.4) & 27.2 (26.1) & 27.0 (25.9) & 26.4 (25.3) & 25.7 (24.2) \bigstrut[b]\\
    \hline
    \end{tabular}
    }
    \tablefoot{
    \tablefoottext{a}{The numbers of exposures in the NUV and $y$ bands are 4 in the WFS and 16 in the DFS.}
    }
\end{table*}

\section{Detectable strongly lensed SNe in CSST}
\label{sec:s3}

In this section, we first give a brief introduction of the CSST survey including its key specifications and the survey strategy under consideration. Another Monte Carlo simulation is then used to estimate the detectable rates of strongly lensed SNe by combining the pregenerated mock catalog and the CSST survey strategy.

\subsection{The China Space Station Telescope}

The CSST is a two-meter space telescope to be launched around 2026. It will co-orbit with the Tiangong Space Station and can dock with the space station for instrument maintenance, upgrades, and refuelling operations regularly or when needed \citep{Zhan2021}. The telescope adopts a Cook-type off-axis three-mirror anastigmat design and will be equipped with five instruments including a survey camera, a terahertz receiver, a multichannel imager, an integral field spectrograph, and a cool-planet imaging coronagraph \citep{Zhan2021}. 

\begin{figure}
       \centering
	\includegraphics[width=0.75\columnwidth]{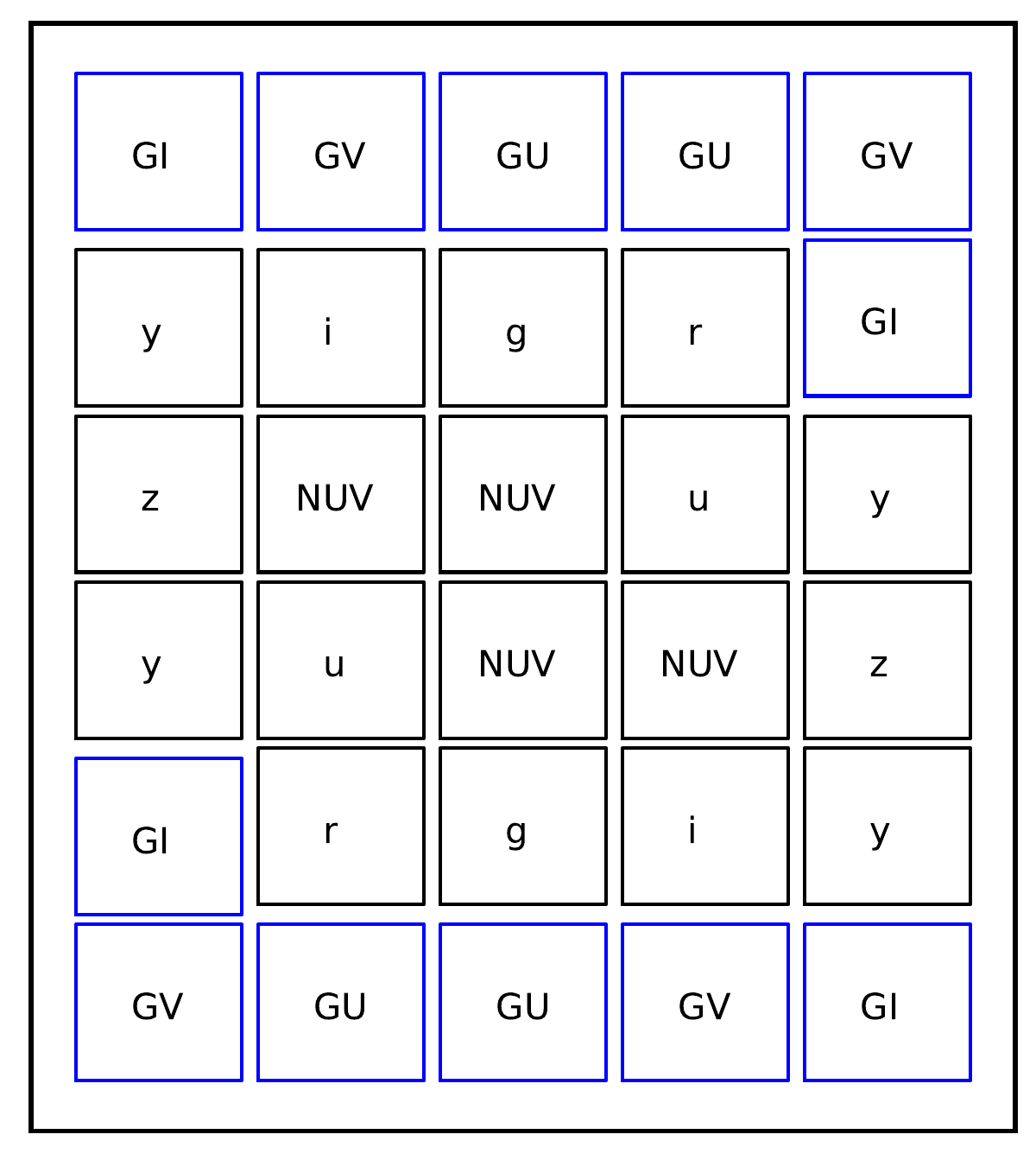}
    \caption{Schematic plot of the focal plane layout of the CSST survey camera. The 18 black squares correspond to detectors designated for multiband imaging observations (in NUV,
    $u$, $g$, $r$, $i$, $z$ and$y$) and the 12 blue squares correspond to detectors designated for slitless spectroscopic observations in GU(255-400 nm), GV(400-620 nm), and GI(620-1000 nm). }
    \label{fig:ccd_plane}
\end{figure}

The primary mission of the CSST 
%$$, known as the Chinese Space Station Optical Survey (CSS-OS),
is to carry out large-scale multiband imaging and slitless spectroscopy surveys simultaneously with the survey camera. The survey camera consists of 30 detectors, each 9K $\times$ 9K, with a pixel size of 0.074 arcsec and a total field of view of $\approx$1.1 deg$^2$. It is placed on the main focal plane and the arrangement of the detectors is indicated in Fig. \ref{fig:ccd_plane}. Of these 30 detectors, 18 will be used for multiband imaging (i.e., NUV, $u$, $g$, $r$, $i$, $z$, and $y$) and the remaining 12 detectors will be used for slitless spectroscopy. Both imaging and spectroscopic observations will be in the wavelength range of 255-1000 nm. For imaging observations, the radius of 80\% encircled energy, $R_{\rm EE80}$, is expected to be no larger than 0\farcs15. The CSST is designed to complete two programs in ten years, a wide-field survey program (WFS) covering 17,500 deg$^2$ and a deep-field survey program (DFS) covering 400 deg$^2$. On average, every sightline in the WFS and DFS footprints will be visited 2 and 8 times in the $u$, $g$, $r$, $i$, and $z$ bands, respectively, and 4 and 16 times in the NUV and $y$ bands. In the $i$ band, the $5 \sigma$ limiting magnitude for point sources after stacking all exposures is 25.9 mag for the WFS and 27.0 mag for the DFS. The limiting magnitudes in the other bands and other key specifications are provided in Table \ref{tab:band_parameter}. The single-visit limiting magnitude is $2.5\mathrm{log}\sqrt{N}$ ($N$ is the number of exposures) shallower than the stacked limiting magnitude. In this work, we focus on estimating the number of strongly lensed SNe from multiband imaging observations by the CSST. 
\begin{figure}
        \centering
	\includegraphics[width=0.85\columnwidth]{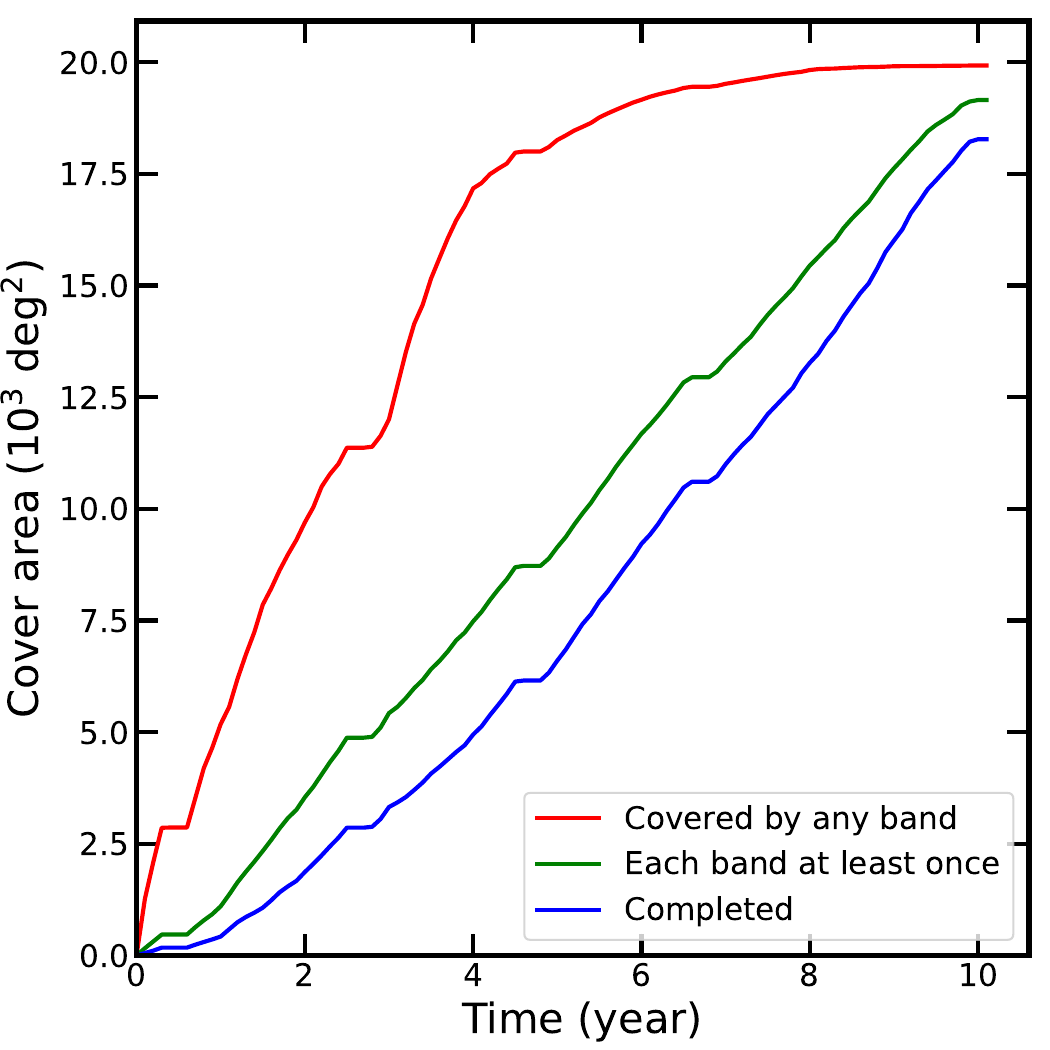}
    \caption{Growth of the sky coverage of the considered survey strategy. The lines correspond to three cases: covered at least once by any band (red), covered at least once by all seven bands (green), and covered by all seven bands to the designed depths (blue).}
    \label{fig:area_time}
\end{figure}

\begin{figure*}
       \centering
	\includegraphics[width=0.7\textwidth]{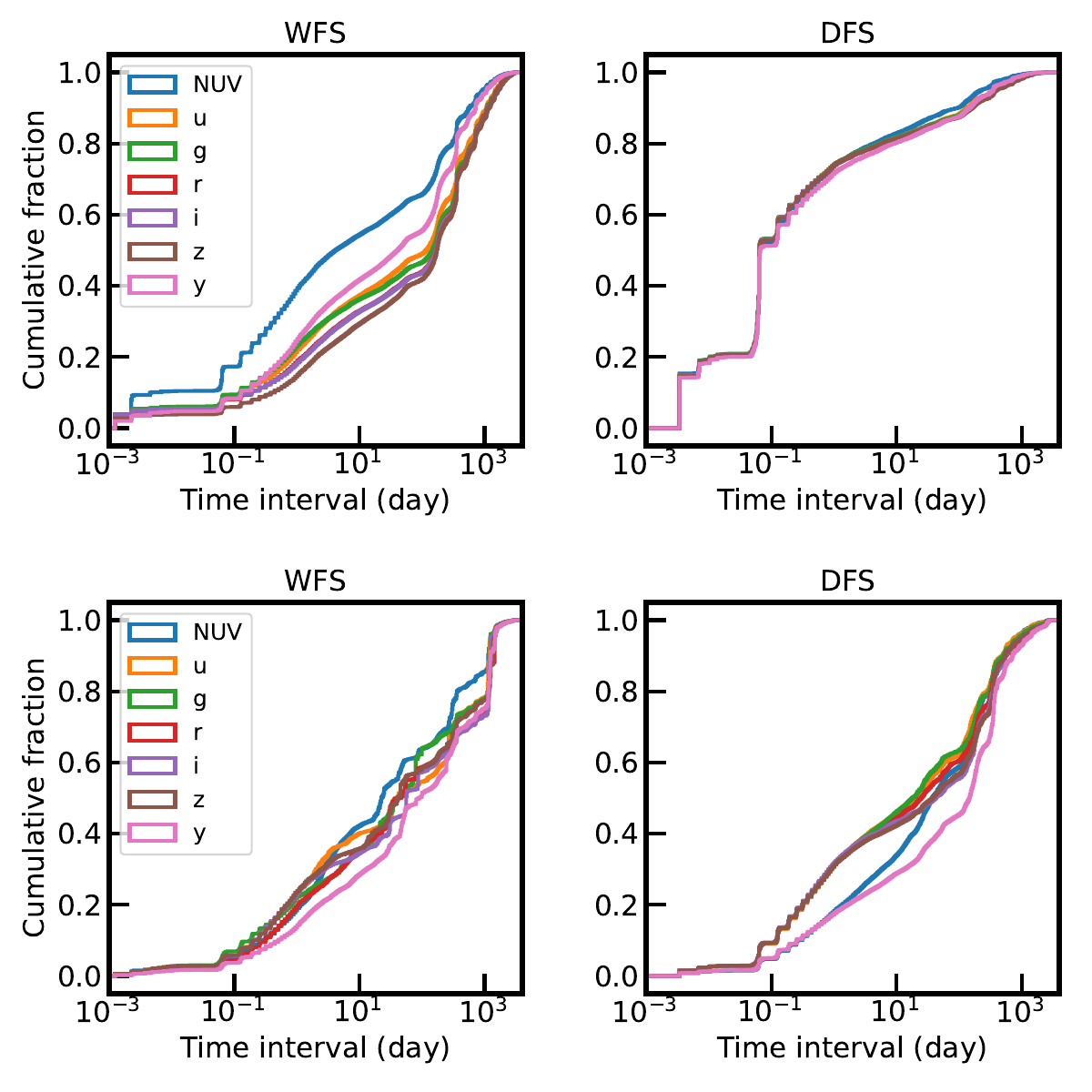}
    \caption{Characteristics of the CSST survey strategy.
    Top: Cumulative distributions of the time interval between two consecutive visits for the WFS and DFS.
    Bottom: Cumulative distributions of the time interval between any two visits for the WFS and DFS.}
    \label{fig:time_interval}
\end{figure*}

The survey strategy of the CSST is essentially a schedule of the telescope pointing, which, at the same time, determines which part of the sky is covered by the survey camera (and the individual detectors) at any given time. The design of the survey strategy needs to take into account technical constraints, such as the solar avoidance angle, the South Atlantic Anomaly standby, regular resupply and maintenance, and so on. In addition, individual scientific goals demand specific adjustments to the survey strategy and authors have explored the impact of the survey strategy on the scientific yields in certain case \citep[e.g.,][]{Fu2023, Yao2024, Liu2024}. In the present work, we consider the latest CSST survey strategy.

The growth of the sky coverage of the considered survey strategy is shown in Fig. \ref{fig:area_time}. Three types of coverage are presented, namely area covered at least once by any band, area covered at least once by all bands, and area reaching the designed depths in all bands. We find that, under this survey strategy, the first type of coverage grows rapidly during the first four years, after which growth slows down significantly. As a result, the planned $\approx$18,000 deg$^2$ footprint will be imaged at least once by at least one band in less than five years. The latter two types of coverage grow roughly linearly with time. The four plateaus in the growth curves indicate the four planned dockings with the Tiangong Space Station. 

We further examined the distributions of the time interval between two consecutive visits and any two visits in each individual band for the WFS and DFS, as shown in Fig. \ref{fig:time_interval}. The median time interval between two consecutive visits is about 5--200 days for the seven bands in the WFS, while for the DFS, the majority of the consecutive visits in each band happen within 1 day. Regarding the cumulative distributions of the time interval between any two visits, they rise earlier in the DFS compared to the WFS, especially in the $u$, $g$, $r$, $i$, and $z$ bands, and the cumulative distributions in the DFS are generally above the cumulative distributions in the WFS between $\approx$0.1--100 days. This indicates that the average time interval between any two visits in the DFS is shorter than that in the WFS.

\begin{figure}
    \centering
	\includegraphics[scale=0.6]{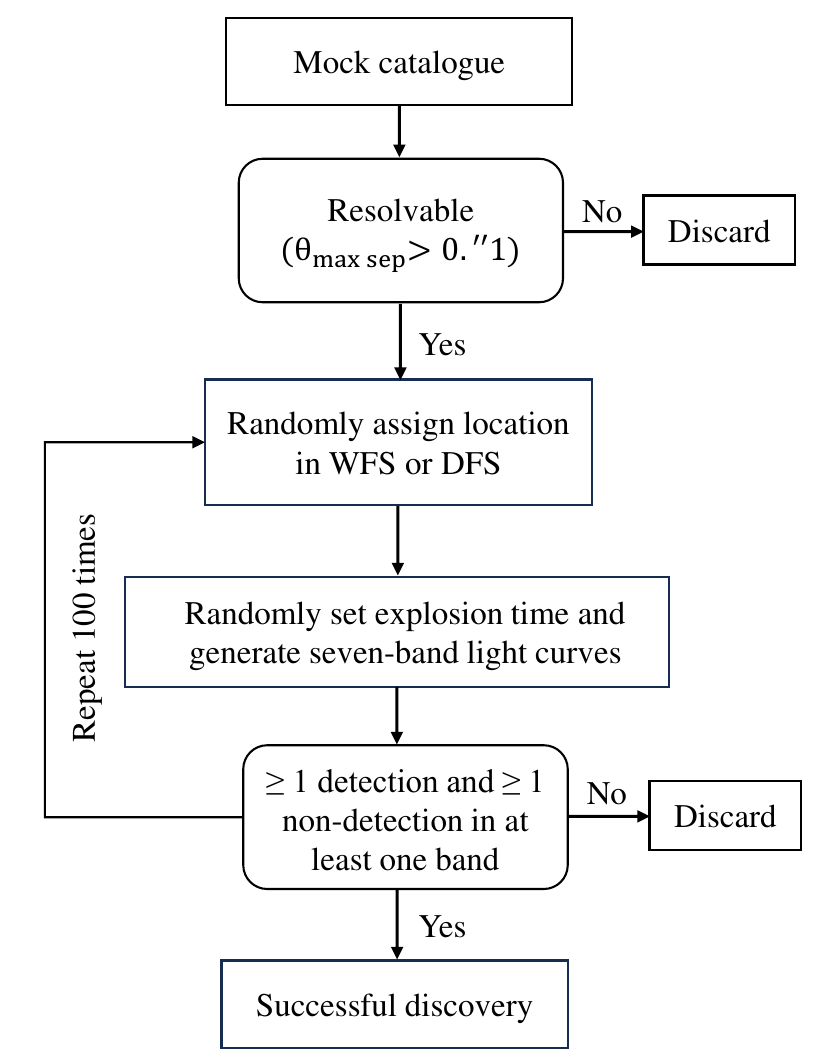}
    \caption{Flowchart of the Monte Carlo simulations for estimating strongly lensed SNe detectable by the CSST.}
    \label{fig:flowchart}
\end{figure}
\begin{table*}
  \centering
  \caption{Forecast results on the number of strongly lensed SNe detectable by the CSST within the ten-year survey duration.}
    \begin{tabular}{clcccccccccc}
    \hline
    \hline
    \multirow{2}[2]{*}{Program } & \multicolumn{1}{c}{\multirow{2}[2]{*}{Condition}} & \multirow{2}[2]{*}{normal Ia} & \multirow{2}[2]{*}{91bg} & \multirow{2}[2]{*}{91T} & \multirow{2}[2]{*}{Ib/c} & \multirow{2}[2]{*}{IIP} & \multirow{2}[2]{*}{IIL} & \multirow{2}[2]{*}{IIn} & \multirow{2}[2]{*}{Ia} & \multirow{2}[2]{*}{CC} & \multirow{2}[2]{*}{Total} \bigstrut[t]\\
      &   &   &   &   &   &   &   &   &  \bigstrut[b]\\
    \hline
    \multirow{2}[1]{*}{WFS} & All & 283.76 & 1.95 & 65.12 & 24.99 & 42.76 & 14.89 & 575.06 & 350.83 &657.70&1008.53 \bigstrut[t]\\
                    & Before-peak & 120.43 & 0.72 & 25.16 & 10.94 & 9.59 & 4.28 & 129.81 & 146.31 &154.62&300.93 \\
    \multicolumn{5}{c}{} &   &   &   &   &  \\
    \multirow{2}[1]{*}{DFS} & All & 15.53 & 0.13 & 3.21 & 1.63 & 2.66 & 1.06 & 27.56 &18.87&32.91& 51.78 \\
                    & Before-peak & 7.20 & 0.07 & 1.45 & 0.78 & 0.73 & 0.35 & 7.51 & 8.72&9.37&18.09 \bigstrut[b]\\
    \hline
    \end{tabular}%
  \label{tab:prediction_result}%
\end{table*}%

\subsection{Rates of detectable strongly lensed SNe in the CSST}
\label{sec:s3.2}

Considering the spatially varying cadences, we performed another set of Monte Carlo simulations to estimate the number of strongly lensed SNe that can be detected by the CSST (Fig.~\ref{fig:flowchart}). Specifically, for any given strongly lensed SN in the mock catalog, (1) we first checked whether its maximum image separation is above 0\farcs1, which is larger than $(2/3) R_{\rm EE80}$ and considered resolvable by the CSST; (2) if yes, we randomly assigned it a projected position on the sky covered by the WFS or DFS;
(3) we randomly set an explosion time between the start and end of the CSST survey and generated light curves for the first arrival SN image (with lensing magnification and Galactic extinction\footnote{We use the \citet{SF2011} recalibration of the \citet{SFD1998} reddening map and the \citet{C1989} extinction law with $R_V=3.1$.} included ) in the seven CSST bands (i.e., NUV, $u$, $g$, $r$, $i$, $z$, and $y$) using the \texttt{SNCosmo} package\footnote{In \texttt{SNCosmo}, every source model has a minimum phase and a maximum phase, outside which the flux of the SN is set to the flux at the minimum or maximum phase (i.e., remaining flat). We think this treatment is unrealistic, and so we assign a constant 40 mag as the magnitude before the minimum phase and extrapolate the light curve beyond the maximum phase based on a linear fit to the SN light curve within rest-frame ten days before the maximum phase.} \citep{Barbary2024}; (4) we then checked the survey schedule to determine whether this lensed SN image would be detected (i.e., above the single-visit limiting magnitude) when visited by the CSST. We consider the selected lensed SN to be successfully discovered if there are at least one detection and one nondetection in at least one (same) band (indicative of a transient source). This requirement also allows a relatively straightforward difference imaging analysis. The above four steps are repeated 100 times for every strongly lensed SN in the mock catalog. Dividing the total number of successful discoveries by ten and multiplying the ratio of the area covered by WFS or DFS to $4\pi$ gives the number of strongly lensed SNe detectable by the WFS or DFS in ten years.

\begin{figure*}
       \centering
	\includegraphics[width=0.96\textwidth]{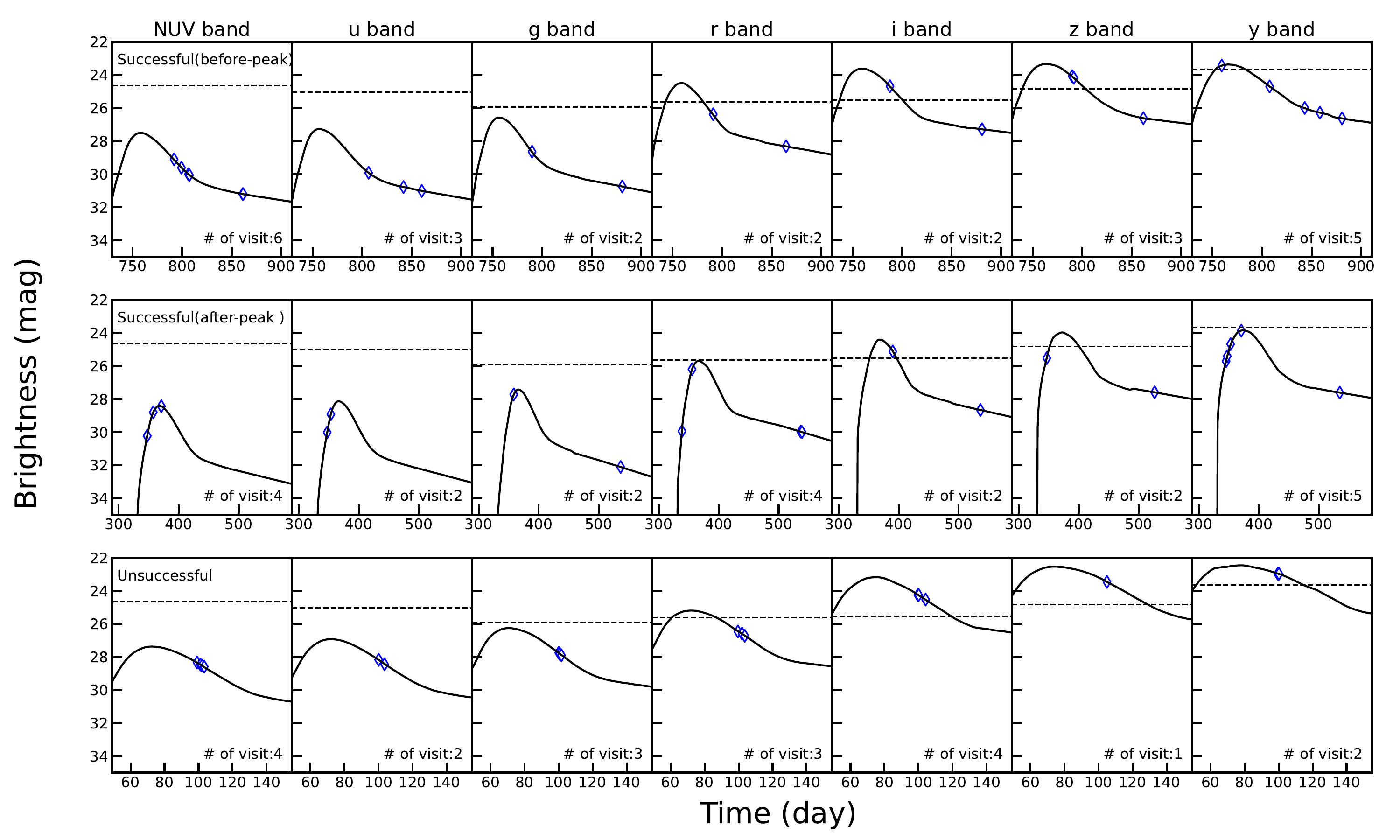}
    \caption{Examples of successful and unsuccessful strongly lensed SNe discoveries in the WFS. From top to bottom, a successful before-peak discovery, a successful after-peak discovery, and an unsuccessful discovery are shown. The seven panels in each row correspond to light curves of the first arrival SN image in the seven CSST bands. The blue symbols indicate the time this system is visited by the CSST. The dashed line in each panel represents the single-visit limiting magnitude in that band.}
    \label{fig:lc_wide}
\end{figure*}
\begin{figure*}
        \centering
	\includegraphics[width=0.96\textwidth]{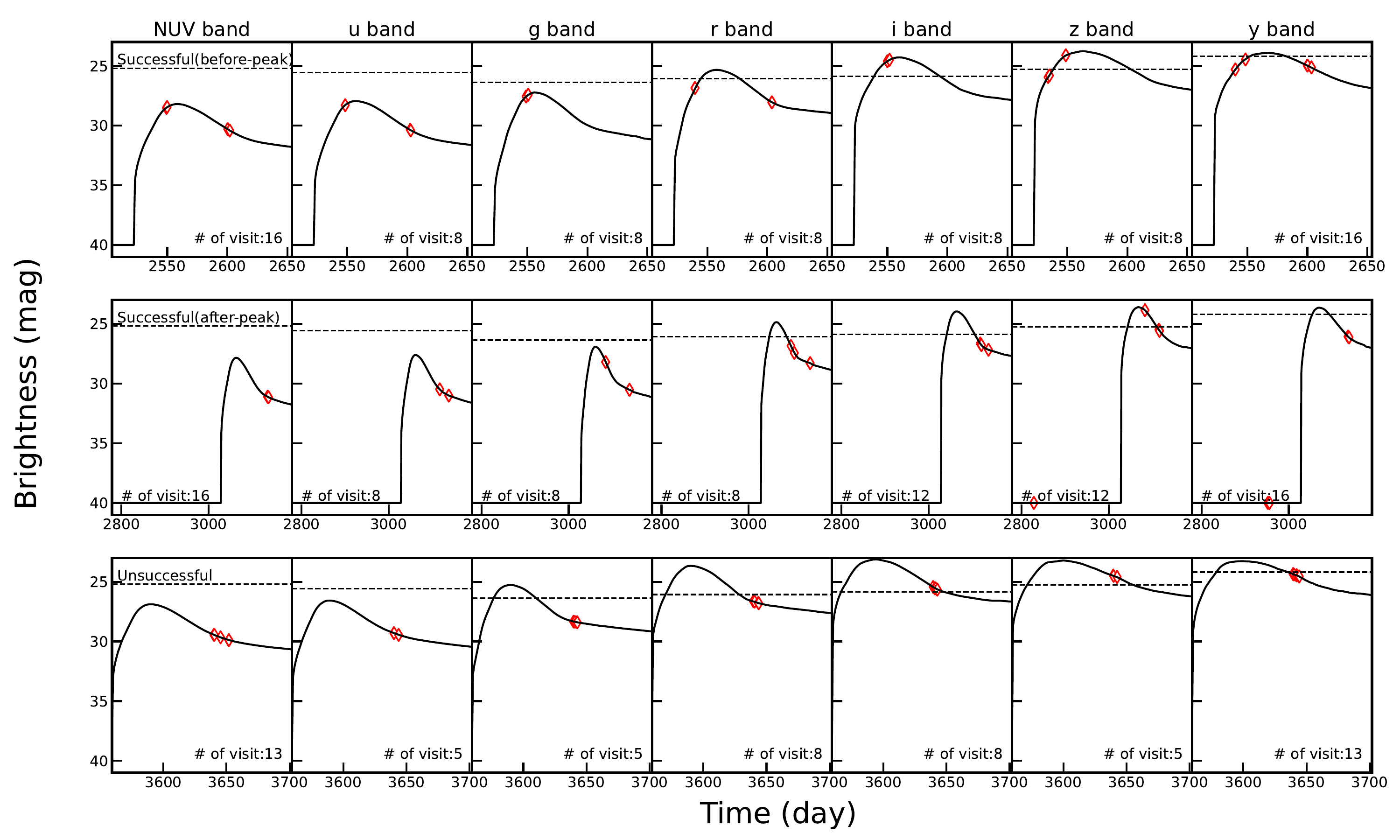}
    \caption{Same as Fig. \ref{fig:lc_wide} but for the DFS.}
    \label{fig:lc_deep}
\end{figure*}

\begin{figure*}
        \centering
	\includegraphics[scale=0.7]{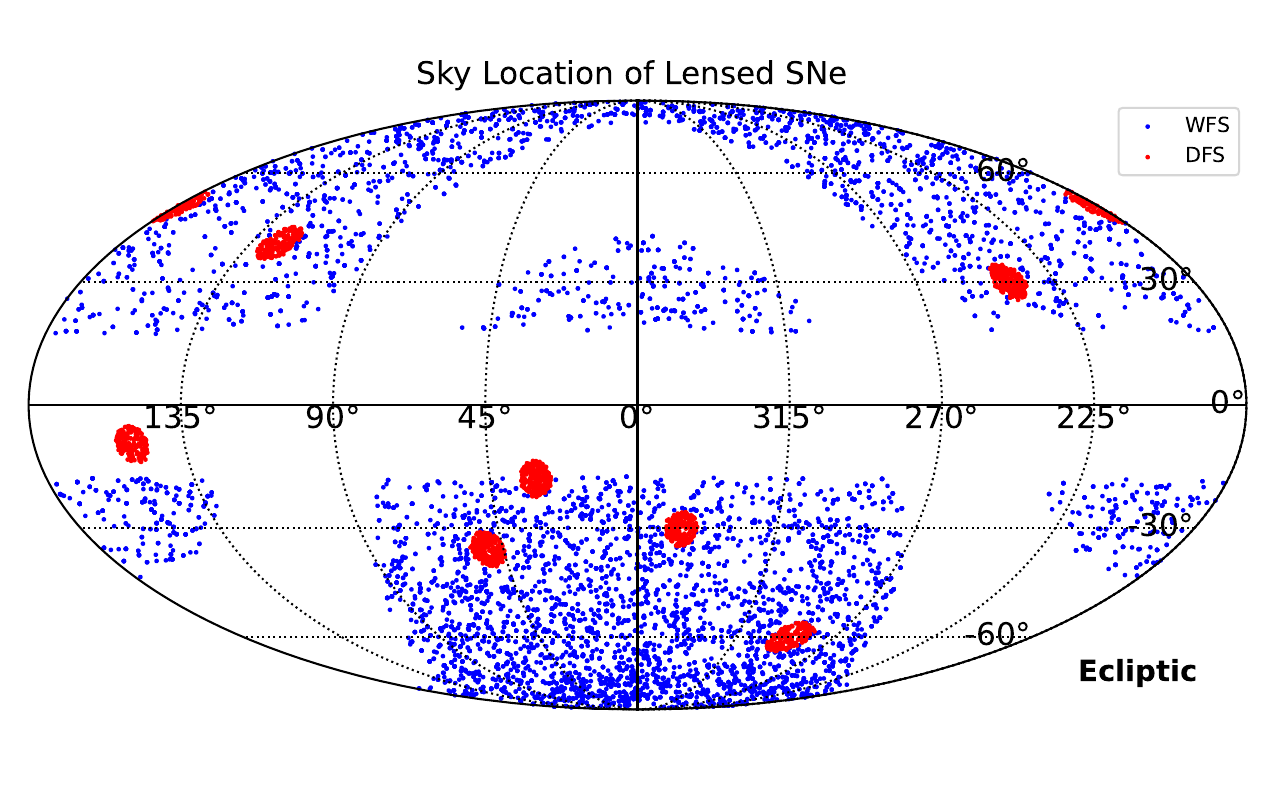}
    \caption{Illustration of the on-sky distribution (in the ecliptic coordinate system) of the strongly lensed SNe detectable in the WFS (blue) and DFS (red). }
    \label{fig:location_SNe}
\end{figure*}

We also separately note down the number of successful discoveries with the first CSST detection happening before the first arrival SN image reaching its peak brightness in the detected band. This type of strongly lensed SN is particularly valuable for time-delay cosmography because more accurate and precise time-delay measurements are expected when the leading image is detected before peak \citep[e.g.,][]{Pierel19}. 

As an illustration, Fig. \ref{fig:lc_wide} and Fig. \ref{fig:lc_deep} give examples of two successful and one unsuccessful strongly lensed SN discovery in the WFS and DFS. Successful discoveries are clearly seen to be those that have at least one detection and one nondetection in at least one band. For the two successful examples in the WFS, the time intervals between two consecutive visits are 0.06-4.1 days and 0.002-181.3 days . For the two successful examples in the DFS, the time intervals are 0.003-6.5 days and 0.003-254.8 days. 

\begin{figure*}
       \centering
	\includegraphics[width=\textwidth]{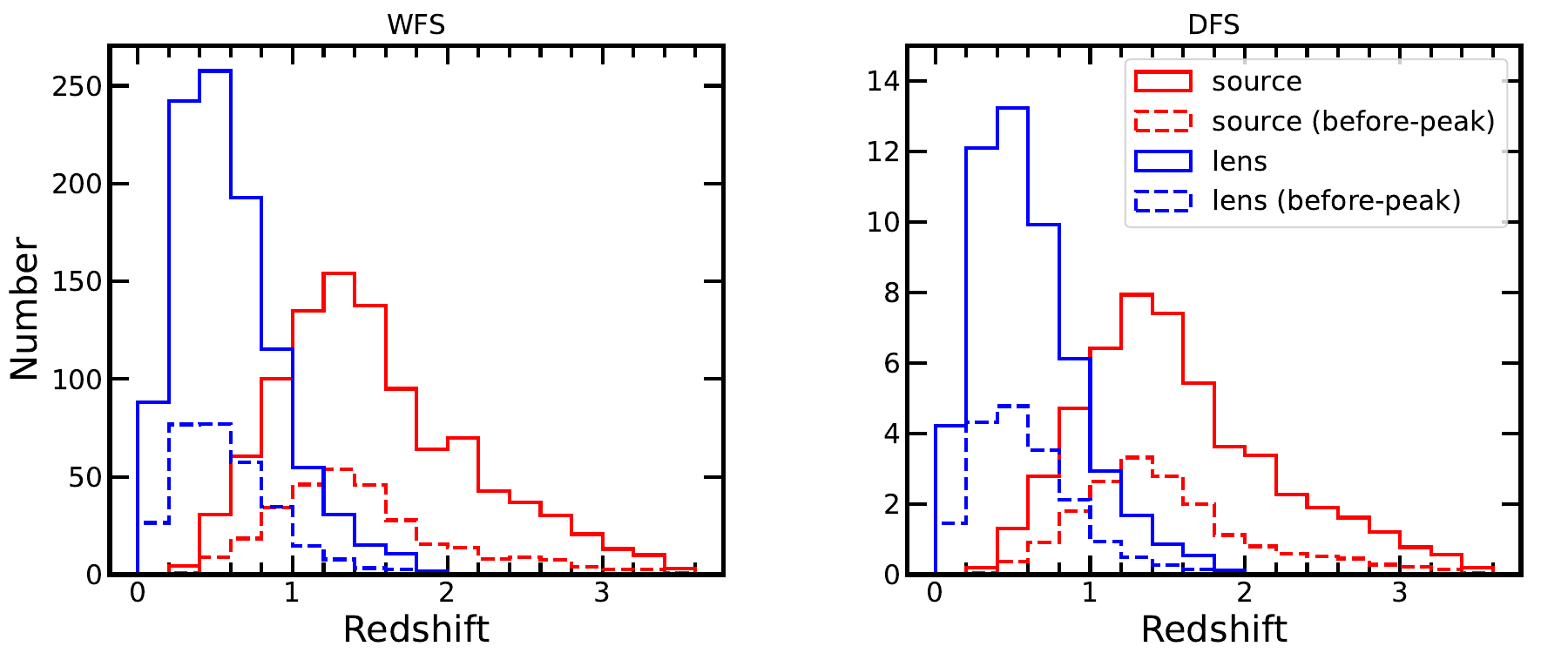}
    \caption{Redshift distributions of the lensing galaxies and lensed SNe of the CSST detectable events in the WFS and DFS.}
    \label{fig:zs_zl_detected}
\end{figure*}
\begin{figure*}
        \centering
	\includegraphics[width=\textwidth]{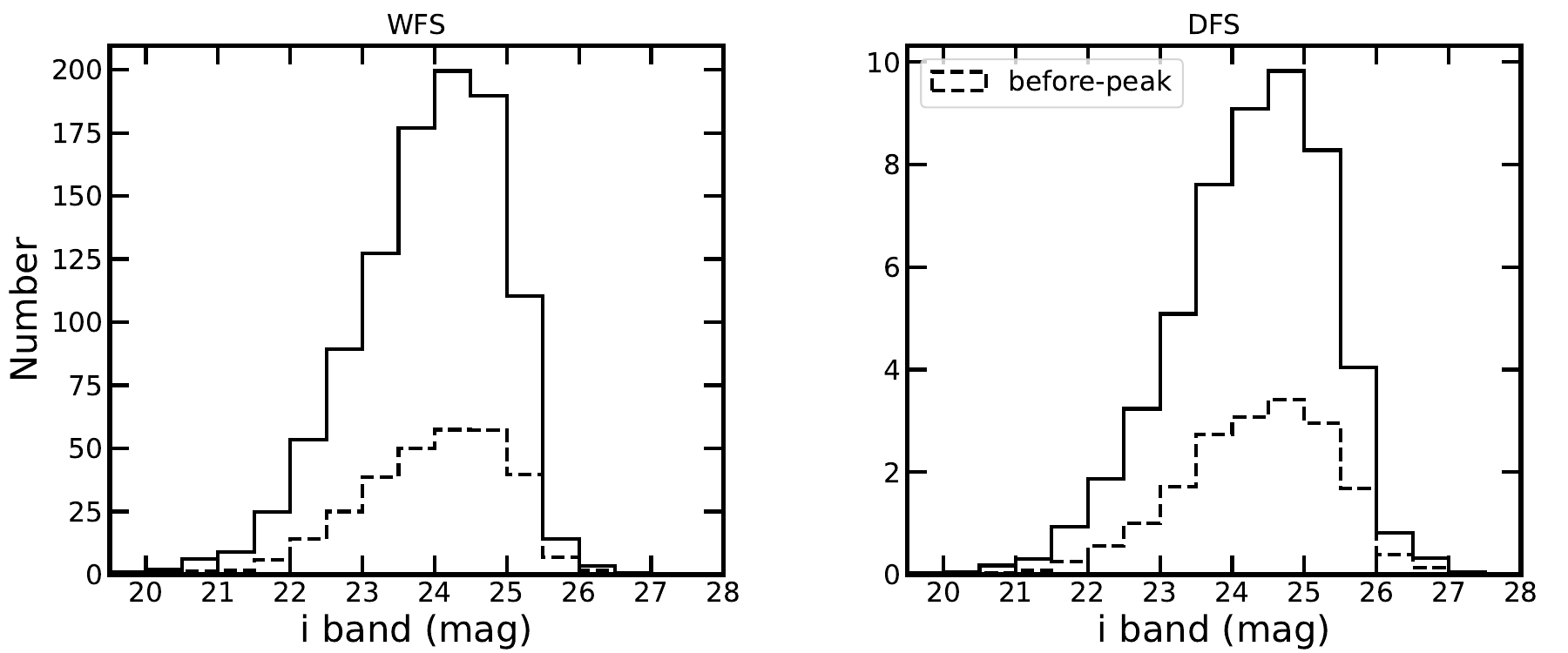}
    \caption{Peak $i$-band apparent magnitude distributions of the lensed SNe of the CSST detectable events in the WFS and DFS.}
    \label{fig:mag_detected}
\end{figure*}

Table \ref{tab:prediction_result} presents our forecast results on the number of strongly lensed SN events detectable by the CSST during the ten-year survey duration. Under the considered survey strategy, we predict that the CSST will detect (detect before peak) in ten years 350.83 (146.31) lensed SNe Ia and 657.70 (154.62) lensed CCSNe in the WFS, and 18.87 (8.72) lensed SNe Ia and 32.91 (9.37) lensed CCSNe in the DFS (Fig.~\ref{fig:location_SNe}). Splitting according to SN subtype, the detectable sources in the WFS (DFS) consist of 28.13\% (30.00\%) normal Ia, 0.19\% (0.24\%) 91bg, 6.46\% (6.20\%) 91T, 2.48\% (3.14\%) Ib/c, 4.24\% (5.13\%) IIP, 1.48\% (2.05\%) IIL, and 57.02\% (53.24\%) IIn. 
%In both WFS and DFS, the fraction of detected lensed Type IIn SN is all largest more than $50\%$.
In terms of the image multiplicity, in the WFS, 94.38\%, 5.54\%, and 0.08\% of the detectable events are doubles, quads, and cusps, respectively. In the DFS, 94.38\%, 5.52\%, and 0.10\% of the detectable events are doubles, quads, and cusps, respectively. The redshift distributions of the lensing galaxies and lensed SNe of the CSST detectable events are shown in Fig. \ref{fig:zs_zl_detected}. The lensing galaxy redshifts peak at $z_l \approx 0.5$ and the lensed SN redshifts peak at $z_s \approx 1.3$ in the WFS and DFS. The peak $i$-band magnitudes of the detectable lensed SNe in the WFS range from 19.5 to 27.0 mag, and the corresponding peak $i$-band magnitude distribution in the DFS reaches $\approx 0.5$ mag fainter, as shown in Fig. \ref{fig:mag_detected}. We note that those lensed SNe with peak $i$-band magnitudes below the WFS or DFS detection limit are still detected because they are above the detection limit(s) in other (presumably) redder band(s).

\begin{figure}
    \centering
    \includegraphics[width=0.9\columnwidth]{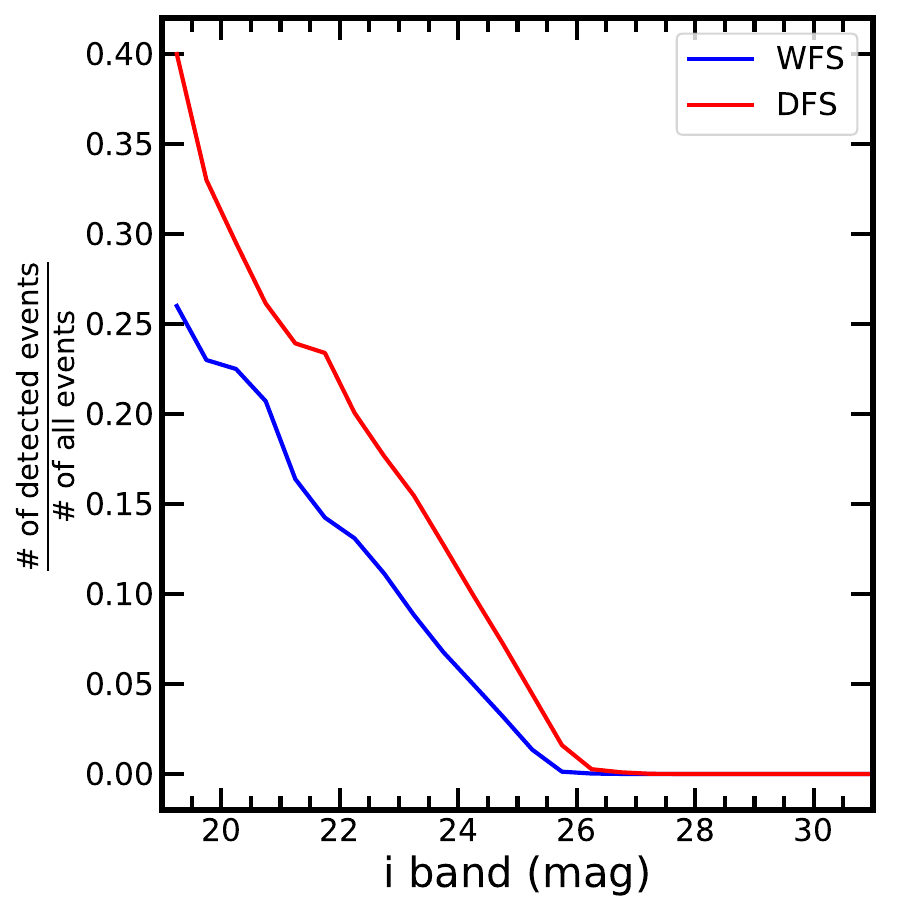}
    \caption{Magnitude dependence of the probabilities of lensed SNe in the mock catalog that are detectable in the WFS (blue) and DFS (red).}
    \label{fig:mag_ratio}
\end{figure}

\section{Discussions}
\label{sec:s4}

Figure \ref{fig:mag_ratio} shows how the probabilities of detection of strongly lensed SNe in the WFS and DFS (after accounting for the sky coverage differences) vary with the peak $i$-band magnitude of the first arrival SN image. We find that these probabilities, or detection efficiencies, decrease with the peak $i$-band magnitude in both WFS and DFS, from $\approx 25$\% ($\approx 40$\%) in the WFS (DFS) on the brightest end to zero beyond $i \approx 25.8$ mag ($\approx 26.3$ mag). We also note that the detection efficiency in the WFS decreases more slowly than in the DFS. 
These trends are understandable because the detection efficiency is mainly related to two timescales: that is, the duration of the first arrival SN image being brighter than the detection limit $\Delta t_1$ and the time span between any two visits in one band $\Delta t_2$. Generally speaking, $\Delta t_1$ decreases with the $i$-band magnitude, which leads to a lower chance of detecting the event. The DFS has better depths than the WFS, and so its detection efficiency drops to zero at a fainter magnitude. On the other hand, at least one nondetection is required for a successful discovery, which implies that, qualitatively, the probability tends to increase with the ratio of $\Delta t_2$ to $\Delta t_1$. As a result, the previous effect, that is, the decrease in the detection efficiency with respect to the peak $i$-band magnitude, will be slowed down. Compared to the WFS, the DFS has shorter $\Delta t_2$ on average (Fig.~\ref{fig:time_interval}), and so its detection efficiency decreases faster. 

The magnitude dependence of the detection efficiencies may be used for rough estimations of other CSST-related events without running a Monte Carlo simulation. For instance, combining the detection efficiencies and the unlensed SN population generated in Sect.~\ref{sec:s2.2}, we predict that approximately 2,700,000 and 138,000 unlensed SNe can be detected in the WFS and DFS in ten years, or equivalently $\approx$15 and 34 unlensed SNe per square degree per year. Among those predicted unlensed SN events, about one-third are SNe Ia. We note that \citet{Lishiyu2023} made a prediction of the number of (unlensed) SNe Ia that can be detected by the CSST from a proposed ultradeep field (UDF) program. This UDF program is supposed to conduct 80 observations of a 10 deg$^2$ field in each of the seven CSST bands in two years. The single-visit exposure time will be 150s in each band, reaching the same depths as the WFS. Assuming a random cadence between 4 to 14 days, these authors found that $\approx$1900 SNe Ia can be detected by the UDF program, or equivalently $\approx$95 per square degree per year. The detectable SNe Ia rate is substantially higher in \citet{Lishiyu2023} mainly because of the high cadence of the UDF program. For a sampling cadence of 4 to 14 days, almost 100\% of the SNe Ia that are beyond the detection limit will be eventually detected. For comparison, the average detection efficiency of unlensed SNe Ia down to $i = 25.8$ mag in the WFS is $\approx$3\%. The two SNe Ia rates are roughly consistent when this difference in detection efficiency is taken into account. 

The uncertainties in our forecast results are dominated by the adopted models and assumptions. In particular, the adopted LFs of individual SN subtypes and their relative abundances, taken from \citet{Li2011}, are based on a volume-limited sample of 175 SNe in the local Universe. Given the small sample size, the uncertainties in the LFs and relative abundances are expected to be significant. In addition, due to the lack of SNe at high redshifts, we assume that the LFs and relative abundances do not evolve within the considered redshift range (0--3.5). We also do not consider the light contamination from the lensing galaxies or the SN host galaxies, which will bias the S/Ns of the lensed SN images.

We emphasize that there is an important difference in the adopted detection criterion from \citet{Oguri2010}. We define detections based on the first arrival image (i.e., using the first arrival image as the reference), while \citet{Oguri2010} used the fainter image as the reference for doubles and the third brightest image for quads and cusps. For doubles, the fainter image is usually the trailing image, while the third brightest image of quads and cusps tends to be the first arrival image. We choose to use the first arrival image as the reference because SNe are transients and we hope to capture all the lensed SN images of a particular lensed SN event (especially doubles) for measuring the time delays. This difference can also explain the difference in the quad fraction. \citet{Oguri2010} found a typical quad fraction of 10-20\% at $i_{\rm lim} \gtrsim 25$ mag, while the quad fraction of the CSST detectable events is $\approx$5.5\%. If requiring the fainter image to be detected for doubles, our simulation also finds a quad fraction of $\approx$25\% at $i_{\rm lim} \sim 25.5$ mag. On the other hand, the relative fractions of doubles, quads, and cusps in our mock catalog (given in Sect. \ref{sec:s2.3}) are considered intrinsic because the adopted magnitude cut there is based on the \textit{unlensed} peak $i$-band apparent magnitude of the source SNe. 
%This is because our catalog contains more doubles as a result of using the brighter image  As we just mentioned, the first arrival image is usually brighter for doubles and the third brightest for quads, which allows us can detect more double images. For example, if we require the second arrival image is detected and adopt a single-visit depth of 25.5 mag in $i$ band, the fraction of doubles decreases from $95\%$ to $74\%$. 

It is highly encouraging to see that CSST will be able to discover strongly lensed SNe on a nearly daily basis. Strongly lensed SNe yields from other ongoing and forthcoming surveys have also been explored in the literature. In particular, \citet{Oguri2010} performed detailed Monte Carlo simulations and estimated that \textit{Rubin} LSST can discover $\approx 130$ strongly lensed SNe (45.7 SNe Ia and 83.9 CCSNe) in 2.5 effective years. The main criteria of a successful discovery in their work include a maximum image separation being above 0\farcs5 and a peak $i$-band magnitude of the selected SN image (the fainter one for doubles and the third brightest one for cusps and quads) being brighter than 22.6 mag. \citet{Goldstein2019} adopted pixel-level simulations to estimate strongly lensed SNe rates in the Zwicky Transient Facility \citep[ZTF,][]{Graham2019} and \textit{Rubin} LSST. These authors suggested that \textit{Rubin} LSST can discover 380.6 strongly lensed SNe per year (54.6 SNe Ia and 326.0 CCSNe) under a nominal observing strategy. \citet{Goldstein2019} did not propose an explanation as to why their LSST prediction was much higher than the prediction in \citet{Oguri2010}. We speculate that this discrepancy is mainly due to two factors. In the forecast of \citet{Goldstein2019}, no limit on the image separation was required. In addition, these latter authors treated the multiple- lensed SN images as a single object when calculating the photometry. As a result, their forecast includes events that have smaller image separations as well as events that are fainter than those considered in \citet{Oguri2010}. \citet{Pierel2021} made a similar forecast for the Roman Space Telescope \citep{Spergel2015}, which suggested approximately $30$ strongly lensed SNe in total. All the forecasts, including ours, suggest that discoveries of large numbers of strongly lensed SNe will soon be made.

Similar to other previous forecast results, we do not consider superluminous supernovae (SLSNe) as the background sources, which belong to a relatively new SN subtype that are typically 10-100 times more luminous than normal SNe. In a study of the absolute magnitude distribution of SNe, \citet{Richardson2002} noted some overluminous SN events ($M_{\rm B} < -21$). \citet{Quimby2007} confirmed the first SLSN discovery, SN 2005ap at $z=0.2832$, which had an unfiltered peak absolute magnitude of -22.7. Nowadays, about $30$ SLSNe are discovered every year \citep[e.g.,][]{Guillochon2017, Chen2023}. Indeed, \citet{Quimby2013b} estimated that the rate of Type I SLSNe (hydrogen-poor) is $32_{-26}^{+77}~\mathrm{Gpc}^{-3}~\mathrm{yr}^{-1} ~h_{71}^{3}$ at a weighted redshift of 0.17 and that the rate of Type II SLSNe (hydrogen-rich) is $151_{-82}^{+151}~\mathrm{Gpc}^{-3}~\mathrm{yr}^{-1}~h_{71}^{3}$ at a weighted redshift of 0.15. A similar SLSN-I rate of $35_{-13}^{+25}~\mathrm{Gpc}^{-3}~\mathrm{yr}^{-1}~h_{70}^{3}$ at $z \leq 0.2$ was reported by \citet{Frohmaier2021}. At higher redshifts, \citet{Prajs2017} measured a volumetric SLSN-I rate of $91_{-36}^{+76}~\mathrm{Gpc}^{-3}~\mathrm{yr}^{-1}$ at $z \sim 1.1$ based on three events. Although these reported SLSN-I and SLSN-II rates are only $\approx$0.1--0.3\% of the used SNe rates at the same redshifts (i.e., Fig.~\ref{fig:sn_rate}), SLSNe are orders of magnitude more luminous than the SNe explored here. We therefore expect that CSST should be able to discover a considerable number of strongly lensed SLSNe as well. Nevertheless, as the SLSNe event rates are currently subject to large uncertainties (due to the small sample size), we chose not to include SLSNe in our simulations.

\section{Conclusions}
\label{sec:s5}

It is well recognized that strongly lensed SNe serve as a promising probe for cosmology and stellar physics \citep[e.g.,][]{Birrer2022b,Huber2022,Grillo2020,Suyu2020,Johansson2021,Bayer2021}. Such applications have been limited by the small sample size (nine discoveries so far). With the advent of next-generation surveys, discoveries of strongly lensed SN events are expected to increase rapidly. In this work, we investigated the capability of detecting strongly lensed SNe with the CSST, a 2m space telescope to be launched around 2026. The primary mission of the CSST includes two programs, WFS and DFS. The WFS will image an area of 17,500 deg$^2$ in seven bands with a single-visit depth of 25.5 mag in the $i$ band. The DFS will image an area of 400 deg$^2$ in the same seven bands with a single-visit depth of 25.9 mag in the $i$ band. 

Using a Monte Carlo simulation, we first constructed a mock catalog of $\approx 6.4 \times 10^4$ strongly lensed SN systems, which correspond to the expected number of strongly lensed SN events per year across the full sky with intrinsic SN peak $i$-band magnitudes down to 31.5 mag. The SN subtypes considered include normal Ia, 91bg, 91T, Ib/c, IIP, IIL, and IIn. The median lens redshift is 0.63, and the median SN redshift is 1.77. Almost 96\% of the mock systems are doubles, 4\% are quads, and 0.07\% are naked cusps. There are also 29 (0.04\%) double source plane lenses. Our mock catalog will be released as a FITS file together with this paper. 

Another Monte Carlo simulation is performed to estimate the number of strongly lensed SNe that can be detected by the CSST by combining the mock catalog and the nominal CSST survey strategy. We find that the WFS and DFS can detect 1008.53 and 51.78 strongly lensed SNe in ten years (Table~\ref{tab:prediction_result}). Among them, 300.93 and 18.09 will be detected in the WFS and DFS before the first-arrival SN images reach peak brightness. %which are expected to deliver more accurate and precise time-delay measurements.
These before-peak detections are expected to deliver more accurate and precise time-delay measurements.
For the events detectable by the CSST, the lensing galaxy redshifts peak at $\approx 0.5$ and the background SN redshifts peak at $\approx 1.3$. For both WFS and DFS, roughly 35\% of the detectable events involve Type Ia SNe.

\begin{acknowledgements}
We would like to thank the anonymous referee for insightful comments that improved the presentation of this work. We are also grateful for the helpful discussions with Dr. Xiaofeng Wang, Dr. Ji-an Jiang, and Mr. Shenzhe Cui.
This work is supported by the China Manned Spaced Project (CMS-CSST-2021-A12, CMS-CSST-2021-A01, CMS-CSST-2021-A03, CMS-CSST-2021-A04, CMS-CSST-2021-A07, CMS-CSST-2021-B01), the National Natural Science Foundation of China (No. 12333001, 11973070, 11333008, 11273061, 11825303, 11673065), the Foundation for Distinguished Young Scholars of Jiangsu Province (No. BK20140050), and the Joint Funds of the National Natural Science Foundation of China (No. U1931210). 

In this study, a cluster is used with the SIMT accelerator made in China. The cluster includes many nodes each containing 2 CPUs and 4 accelerators. The accelerator adopts a GPU-like architecture consisting of a 16GB HBM2 device memory and many compute units. Accelerators connected to CPUs with PCI-E, the peak bandwidth of the data transcription between main memory and device memory is 16GB/s.
\end{acknowledgements}

\bibliographystyle{aa}
\bibliography{ref}

\begin{appendix}
\section{Description of the mock catalog}
\begin{table*}
	\caption{Format of the mock catalog FITS file.}
	\label{tab:catalog}
	\centering
        \scalebox{0.95}{
	\begin{tabular}{lll}
		\hline
            \hline
		   Column &  Name & Description\\
		\hline
		   1 & amp\_shear & the strength of external shear\\
              2 & pa\_shear & the orientation of external shear, in degree\\
              3 & lambda\_q & the dynamical normalization parameter \\
              4 & q\_SIE    & the axial ratio of SIE\\
              5 & v\_disp   & the velocity dispersion of lensing galaxy, in km / s\\
              6 & lens\_redshift & the redshift of lensing galaxy\\
              7 & sn\_type  & the type of supernova,1-7 represent normal Ia, 91bg, 91T, Ib/C, IIP, IIL, IIn, respectively\\
              8 & source\_redshift & the redshift of supernova\\
              9 & source\_xlocation & the x-coordinate of supernova, in arcsec\\
              10 & source\_ylocation & the y-coordinate of supernova, in arcsec\\
              11 & absolute\_mag\_R\_band\_vega & the peak absolute magnitude of supernova in R band in Vega system\\
              12 & apparent\_mag\_i\_band   & the intrinsic peak apparent magnitude of supernova in $i$ band in AB system\\ 
              13 & image\_sep & the max image separation between lensed images, in arcsec\\
              14 & image\_xlocation & the x-coordinate of lensed images, in arcsec\\
              15 & image\_ylocation & the y-coordinate of lensed images, in arcsec\\
              16 & kappa & the covergence of SIE\\
              17 & gamma & the shear of SIE\\
              18 & magnification & the magnification of lensed images\\
              19 & time\_delay & the time delay of lensed images\\
              20 & apparent\_mag\_first\_arrival\_i\_band & the peak apparent magnitude of the first arrival image in i band in AB system\\
              21 & image\_number & the number of lensed images\\
		\hline
	\end{tabular}
          }
\end{table*}

\end{appendix}

\end{document}